\documentclass[submission]{eptcs}
\providecommand{\event}{DICE-FOPARA 2017} % Name of the event you are submitting to
\usepackage{underscore}           % Only needed if you use pdflatex.

\usepackage[utf8x]{inputenc}% WICHTIG! Für rôle etc.
\usepackage[T1]{fontenc}

\usepackage{amsmath,amssymb,amsthm,amsfonts,stmaryrd}
\usepackage{enumerate}

\usepackage{code}
\usepackage{style}
\usepackage{paper}
\usepackage{subcaption}
\usepackage{cleveref}

\newcommand\MS[1]{}
\newcommand\MA[1]{}
\newenvironment{framed}[0]{\begin{boxedminipage}{\linewidth}\vspace{-1mm}}{\vspace{-2mm}\end{boxedminipage}\vspace{-1mm}}
\usepackage{boxedminipage}

\newtheorem{definition}{Definition}
\DeclareMathAlphabet{\mathcal}{OMS}{cmsy}{m}{n}

\title{GUBS Upper Bound Solver \\ \Large{(Extended Abstract)}%
  \thanks{This research is partly supported by DARPA/AFRL contract number FA8750-17-C-088.}}
\author{%
Martin Avanzini \qquad\qquad Michael Schaper
\institute{Department of Computer Science\\University of Innsbruck, Austria}
\email{\{martin.avanzini, michael.schaper\}@uibk.ac.at}
}
\def\titlerunning{GUBS}
\def\authorrunning{M. Avanzini \& M. Schaper}
\begin{document}
\maketitle

%%%%%%%%%%%%%%%%%%%%%%%%%%%%%%%%%%%%%%%%%%%%%%%%%%%%%%%%%%%%%%%%%%%%%%%%%%%%%%

\begin{abstract}
In this extended abstract we present the \gubs{} Upper Bound Solver. 
\gubs{} is a dedicated constraint solver over the naturals for inequalities formed over uninterpreted function symbols and standard arithmetic operations.
\gubs{} now forms the backbone of \hosa{}, a tool for analysing space and time complexity of higher-order functional programs automatically.
We give insights about the implementation and report different case studies.
\end{abstract}

\section{Introduction}
\label{s:introduction}

Synthesizing functions that obey certain constraints in the form of (in)equalities is a fundamental task in program analysis.
For instance, in the context of termination and complexity analysis.
To this end, the program verification community predominantly adopts techniques either based on LP solvers~\cite{PR04}, 
or dedicated tools like \ttt{PUBS}~\cite{AAGP:SAS:08}. The former approach is usually restricted to the synthesis of linear functions.
The latter approach is based on solving recurrence relations and particularly useful for the synthesis 
of ranking functions in the context of imperative programs. 
Recurrence relations are limited in scope however, for instance, function composition cannot be directly expressed.

To overcome these restrictions, we have developed the \emph{\gubs{} Upper Bound Solver} (\gubs\ for short). 
Given a set of inequalities over arithmetical expressions and uninterpreted function symbols, \gubs\ tries to find a model in $\N$,
i.e., an interpretation of all the function symbols that make the given inequalities hold.
This tool is currently used in our inference machinery for sized-types\footnote{See \url{http://cl-informatik.uibk.ac.at/users/zini/software/hosa}.} 
and, experimentally, in our complexity tool \tct\footnote{See \url{http://cl-informatik.uibk.ac.at/software/tct}.} for the synthesis of specific ranking function. 
\gubs\ itself is heavily inspired by methods developed for synthesising linear and non-linear interpretations 
in the context of rewriting. The main novel aspect of \gubs\ is the modular approach it rests upon, 
inspired by the framework underlying \tct.
To date, it incorporates an adaption of the synthesis technique from~\cite{FGMSTZ07} which reduces polynomial inequalities to 
SMT (with respect to the theory of \emph{quantifier free non-linear integer arithmetic}), various syntactic simplification techniques and a per-SCC analysis.
\gubs\ is open source and available from
\begin{center}
\url{https://github.com/ComputationWithBoundedResources/gubs}\tpkt
\end{center}

Although developed foremost for the complexity analysis of (higher-order) rewrite systems, 
we are convinced that \gubs\ has potential applications outside rewriting.
E.g.,\ for the synthesis of ranking functions, the inference of certain dependent type systems such as the one of Dal Lago and Petit~\cite{LP:POPL:13}, 
the complexity analysis of concurrent programs as in \cite{GBCLP:FACS:15}, 
or various systems developed by the implicit computational complexity community (we name just~\cite{BMM:TCS:11}).

In the following we briefly outline our tool \gubs. More specific, we introduce the problem tackled by \gubs\ formally, 
we outline the central synthesis techniques and report on the experience that we have collected so far.

%%% Local Variables:
%%% mode: latex
%%% TeX-master: "paper"
%%% End:

\section{Constraint System over the Naturals}\label{s:system}

Let $\VS$ denote a countably infinite set of \emph{variables}, and 
let $\FS$ denote a \emph{signature}, i.e., a set of \emph{function symbols}, 
disjoint from $\VS$. 
Each function symbol $\fun{f} \in \FS$ is equipped with a natural number $\ar{\fun{f}}$, its \emph{arity}.
We use $x,y,\dots$ to denote variables, whereas $\fun{f},\fun{g},\dots$ denote function symbols.
The set of \emph{arithmetical terms} $\TERMS$ over function symbols $\FS$ and variables $\VS$ 
is generated inductively from $x \in \VS$, $\fun{f} \in \FS$, $n \in \N$ and pre-defined arithmetical operations 
${\oplus} \in \{\PLUS,\MUL,\MAX\}$ according to the following grammar:
\[
  \termone,\termtwo \bnfdef x \mid n \mid \termone \oplus \termtwo \mid \fun{f}(\termone_1,\dots,\termone_{\ar{\fun{f}}}) \tpkt
\]
A \emph{constraint system} $\CS$ (over $\N$) is a finite set of \emph{constraints} $\termone \cs \termtwo$. 

Informally, a constraint system $\CS$ is \emph{satisfiable} if we can interpret each symbol $\fun{f} \in \FS$ with a
function $\inter{\fun{f}} \ofdom \N^{\ar{\fun{f}}} \to \N$ such that all constraints in $\CS$ hold. Here, constraints and arithmetical operations 
are interpreted in the natural way. Consider for instance the constraint system $\CS_1$ consisting of the following two constraints:
\begin{align*}
  \fun{r}(\fun{n},y) & \cs 1 & \fun{r}(\fun{c}(x,y),z) & \cs 1 + \fun{r}(x,\fun{c}(y,z)) \tpkt
\end{align*}
This system is satisfiable, for instance, 
by taking $\inter[\II]{\fun{n}} = 1$, $\inter[\II]{\fun{r}}(x,y) = x$ and $\inter[\II]{\fun{c}}(x,y) = y + 1$, 
as we have 
\[
\inter[\II]{\fun{r}}(\inter[\II]{\fun{n}},y) = 1 \cs 1  \quad\text{and}\quad \inter[\II]{\fun{r}}(\inter[\II]{\fun{c}}(x,y),z) = y + 1 \cs 1 + y = 1 + \fun{r}(y,\fun{c}(x,z)) \tpkt
\]

More formally, an \emph{interpretation} $\II$ over symbols $\FS$ (into the naturals) is a mapping that assigns to each symbol $\fun{f} \in \FS$ 
a function $\inter{\fun{f}} \ofdom \N^{\ar{\fun{f}}} \to \N$. 
Let ${\oplus_\N} \ofdom \N^2 \to \N$ for ${\oplus} \in \{{\PLUS},{\MUL},{\MAX}\}$ denote addition, multiplication and the maximum function on natural numbers, respectively.
The interpretation $\im[\alpha]{t}$ of a term $t \in \II$ with respect to $\II$ and variable assignment $\alpha \colon \VS \to \N$ is then defined in the expected way:
\[
  \im[\alpha]{\termone} =
\begin{cases}
  \alpha(\termone)                                                         & \text{ if } \termone \in \VS\tkom\\
  \termone                                                                 & \text{ if } \termone \in \N\tkom\\
  % \termone_1 \oplus_\N \termone_2                                          & \text{ if } \termone = \termone_1 \oplus \termone_2 \in \N \text{ and } {\oplus} \in \{\PLUS,\MUL,\MAX\}\tkom\\
  \im[\alpha]{\termone_1} \oplus_\N \im[\alpha]{\termone_2}                & \text{ if } \termone = \termone_1 \oplus \termone_2 \text{ and } {\oplus} \in \{\PLUS,\MUL,\MAX\}\tkom\\
  \inter{\fun{f}}(\im[\alpha]{\termone_1},\ldots,\im[\alpha]{\termone_n}), & \text{ if }\termone=\fun{f}(\termone_1,\ldots,\termone_n)\text{ and } \fun{f} \in \FS\tpkt
\end{cases}
\]
We say that $\II$ is a \emph{model} of a constraint system $\CS$, in notation $\II \models \CS$, if $\im[\alpha]{\termone} \geqslant_\NN \im[\alpha]{\termtwo}$ holds for all assignments $\alpha$ 
and constraints $\termone \cs \termtwo \in \CS$. 
For instance, we have $\II \models \CS_1$ for the constraint system $\CS_1$ and interpretation $\II$ depicted above. 
With our tool \gubs, we give a sound, but necessarily incomplete procedure to the following undecidable problem.
\begin{definition}[Model Synthesis]
  Given a constraint system $\CS$, the \emph{model synthesis problem} asks for an interpretation $\II$ with 
  $\II \models \CS$.
\end{definition}

%%% Local Variables:
%%% mode: latex
%%% TeX-master: "paper"
%%% End:

\section{Implementation}
\label{s:implementation}

\gubs\ is written in the functional programming language \ttt{Haskell}. The source 
contains approximately 2000 lines of code, spread over 25 modules. 
\gubs\ comes along as a stand-alone executable as well as a \ttt{Haskell} library. 
For usage information and installation instructions, we kindly refer the reader to the homepage of \gubs.
Here, we just provide a short outline of the central methods implemented in \gubs{}.

\paragraph{Synthesis of Models via SMT.}

% Given a constraint system $\CS$ as defined in Section~\ref{s:system}, the problem of finding a model $\II$ is undecidable.
% Approximations for related systems have been investigated before.
Conceptually, we follow the method presented in~\cite{FGMSTZ07}. 
% Inequalities over unknown polynomials are reduce to formulas over the theory \emph{quantifier free non-linear integer arithmetic}.
In this approach, each $k$-ary symbol $\fun{f} \in \FS$ is associated with a \emph{template max-polynomial}, i.e., 
an expression over $k$ variables and connectives $\{{\PLUS},{\MUL},{\MAX}\}$, and undetermined coefficient variables $\vec{\m{c}}$. 
For instance, a linear template for a binary symbol $\fun{f}$ employed by \gubs\ is
\[
  \inter[\AS]{\fun{f}}(x,y) = \max(\m{c_1} \cdot x + \m{c_2} \cdot y + \m{c}_3, \m{d_1} \cdot x + \m{d_2} \cdot y + \m{c}_3 ) \tpkt
\]
% This then allows the interpretation of terms $\termone$ as max-polynomials $\im[][\AS]{\termone}$. 
% Finding a model then amounts to finding suitable coefficient $\vec{n} \in \N$ for the occurring coefficient variables $\vec{c}$ so that 
% $\im[\AS]{\termone}\{ \vec{\m{c}} \slash \vec{n} \} \geq \im[\AS]{\termtwo}\{ \vec{\m{c}} \slash \vec{n} \}$ holds. 
% The concrete model is then given by $\AS$, with coefficient variables $\vec{\m{c}}$ substitute by $\vec{n}$. 
To find a concrete model based on these templates, we search for a solution to
$\exists \vec{\m{c}}.\ \bigand_{\termone \cs \termtwo \in \CS} \forall \vec{x} \in \VS.\ \im[\AS]{\termone} \geq \im[\AS]{\termtwo}$, in two steps. 
First, we eliminate $\max$ according to the following rules.
\begin{align*}
  e \geq C[\max(f_1,f_2)] & \quad\Rightarrow\quad e \geq C[f_1] \land e \geq C[f_2] \tkom\\
  C[\max(e_1,e_2)] \geq f & \quad\Rightarrow\quad C[e_1] \geq f \lor C[e_2] \geq f \tpkt
\end{align*}
Notice that this elimination procedure is sound as our max-polynomial algebra allows the formation of weakly monotone expressions only. 
Once all occurrences of $\max$ are eliminated, we reduce the resulting formula to \emph{diophantine constraints} over the coefficient variables $\vec{\m{c}}$, 
via the so called \emph{absolute positiveness check}, see also~\cite{FGMSTZ07}. The diophantine constraints are then given to an SMT-solver that supports quantifier-free non-linear integer arithmetic, 
from its assignment and the initially fixed templates \gubs\ then computes concrete interpretations. 
To get more precise bounds, \gubs\ \emph{minimises} the obtained model by making use of the incremental features of current SMT-solvers, 
essentially by putting additional constraints on coefficients. 

\emph{Limitations}:
The main limitation of this approach is that the shape of interpretations is fixed to that of the templates, noteworthy, the degree of the interpretation is fixed in advance. 
As the complexity of the absolute positiveness check depends not only on the size of the given constraint system but to a significant extend also on the degree of interpretation functions, 
our implementation searches iteratively for interpretations of increasing degree.

Also notice that our max-elimination procedure is incomplete, for instance, it cannot deal with the constraint $\max(2x,2y) \geqslant x + y$, which is reduced to $2x \geqslant x + y \lor 2y \geqslant x + y$. 
In contrast, in~\cite{FGMSTZ:RTA:08} a complete procedure is proposed. However, our experimental assessment concluded that this encoding introduces too many auxiliary variables, 
which turned out as a significant bottleneck of the overall procedure.  

\paragraph{Separate SCC Analysis.}
Synthesis of models via SMT gets impractical on large constraint systems. To overcome this, \gubs\ divides the given constraint system $\CS$ into 
its \emph{strongly connected components} (\emph{SCCs} for short) $\CS_1,\dots,\CS_n$, topologically sorted, and finds a model for each SCC $\CS_i$ iteratively. 
Here, the underlying \emph{call graph} is formed as follows. The nodes are given by the constraints in $\CS$. 
Let $\termone_1 \cs \termtwo_1$ to $\termone_2 \cs \termtwo_2$ be two constraints in $\CS$, 
where wlog.\@ $\termone_1 = C[\fun{f_1}(\vec{\termone_1}),\dots,\fun{f_n}(\vec{\termone_n})]$ for a context $C$ without function symbols.
Then there is an edge from $\termone_1 \cs \termtwo_1$ to $\termone_2 \cs \termtwo_2$
if any of the symbols occurring in $\vec{\termone_1},\dots,\vec{\termone_n},\termtwo_1$ occurs in $\termone_2$.
The intuition is that once we have found a model for all the successors of $\termone_1 \cs \termtwo_1$, we can interpret 
the terms $\vec{\termone_i}$ and $\termtwo_1$ within this model. We can then extend this model by finding a suitable interpretation for 
$\fun{f_1},\dots,\fun{f_n}$, thereby obtaining a model that satisfies $\termone_1 \cs \termtwo_1$.

\paragraph{Syntactical Simplifications.}
We apply a series of syntactical complete simplifications that are fast and may reduce the search domain for the SMT solver.
For example, \emph{instantiation} substitutes all variables that occur only on the left-hand side of a constraint with $0$.
This reduces the size of the generated abstract polynomials.
\emph{Elimination} fixes the interpretation of function symbols that occur only on the right-hand side of the constraint system to $0$.
This reduces the domain of the interpretation.
\emph{Propagation} performs a restricted form of inlining to simplify the constraint system.
For instance, given a constraint $\fun{f}(x,y) \geqslant x+y$, if $\fun{f}$ does not occur on the left-hand side of a different constraint we can fix the model of $\fun{f}$ and substitute all $\fun{f}$ occurring on the right-hand side of the constraint system.
% \emph{Propagation} is used when $\termone_1 = C[\fun{f_1}(\vec{\termone_1}),\dots,\fun{f_n}(\vec{\termone_n})]

% \MA{divide and conquer (splitting $\CS$) desirable; incomplete (with example?)}
% \MA{SCC analysis => informal def of arc in graph => bottom up}
% \MA{minimisation important}

%%% Local Variables:
%%% mode: latex
%%% TeX-master: "paper"
%%% End:

\section{Case Studies}
\label{s:casestudies}
In this section we briefly outline our experience collected so far, in 
the two contexts where \gubs\ is currently employed.

\paragraph*{Synthesis of Sized-Types.}

\begin{figure}[t]
\begin{subfigure}[b]{\textwidth}
\begin{framed}
  \begin{minipage}{\linewidth}
\centering
\begin{lstlisting}[style=ocaml,basicstyle=\scriptsize\ttfamily,emph={i,j,k,l,ijk,ij},emph={[2] append,map,prependAll,f1,f3,f2,f5,f4,f6}]
       map : $\forall$ijk.($\forall$ l.L[l](a) -> L[f4(l,i)](a)) ->  L[k](L[j](a)) -> L[f6(i,j,k)](L[f5(i,j,k)](a))
    append : $\forall$ij.L[i](a) -> L[j](a) -> L[f1(i,j)](a)
prependAll : $\forall$ijk.L[i](a) ->  L[k](L[j](a)) -> L[f3(i,j,k)](L[f2(i,j,k)](a))
\end{lstlisting}
\end{minipage}
\end{framed}
\caption{Template sized-types assigned by \protect\hosa\ to the main function $\fun{prependAll}$ and auxiliary functions.}\label{fig:hosa:templates}
\vspace{2mm}
\end{subfigure}

\begin{subfigure}[b]{\linewidth}
\begin{framed}
\begin{minipage}{0.49\linewidth}
\begin{lstlisting}[emph={[2] f1,f2,f3,f4,f5,f6,f71,f72,f78,f71,f72,f73,f74,f75,f76,f77,f78,f79,f80,f81,f82,f83,f84,f85,f86,f87},style=cs,emph={x,y,z,w}]
(>= (f1 0 (var x)) (var x))
(>= (f1 (+ (var x) 1) (var y)) 
    (+ (f78 (var y) (var x)) 1))
(>= (f2 (var x) (var y) (var z)) 
    (f5 (f73 (var x)) (f75 (var y)) (f74 (var z))))
(>= (f3 (var x) (var y) (var z)) 
    (f6 (f73 (var x)) (f75 (var y)) (f74 (var z))))
(>= (f4 (var x) (f73 (var y))) 
    (f1 (f71 (var y)) (f72 (var x))))
(>= (f4 (var x) (f82 (var y))) 
    (f4 (f81 (var x)) (var y)))
(>= (f5 (var x) (var y) (+ (var z) 1)) 
    (f80 (var x) (var y) (var z) (var w)))
(>= (f5 (var x) (var y) 0) (f86))
(>= (f6 (var x) (var y) 0) 0)
(>= (f6 (var x) (var y) (+ (var z) 1)) 
    (+ (f85 (var x) (var y) (var z) (var w)) 1))
\end{lstlisting}
\hfill
\end{minipage}
\begin{minipage}{0.45\linewidth}
\begin{lstlisting}[emph={[2] f1,f2,f3,f4,f5,f6,f71,f72,f78,f71,f72,f73,f74,f75,f76,f77,f78,f79,f80,f81,f82,f83,f84,f85,f86,f87},style=cs,emph={x,y,z,w}]
(>= (f71 (var x)) (var x))
(>= (f72 (var x)) (var x))
(>= (f74 (var x)) (var x))
(>= (f75 (var x)) (var x))
(>= (f76 (var x)) (var x))
(>= (f77 (var x)) (var x))
(>= (f78 (var x) (var y)) 
    (f1 (f76 (var y)) (f77 (var x))))
(>= (f79 (var x)) (var x))
(>= (f80 (var w) (var x) (var y) (var z)) 
    (f4 (f79 (var x)) (var w)))
(>= (f80 (var w) (var x) (var y) (var z)) 
    (f5 (f82 (var w)) (f84 (var x)) (f83 (var y))))
(>= (f81 (var x)) (var x))
(>= (f83 (var x)) (var x))
(>= (f84 (var x)) (var x))
(>= (f85 (var w) (var x) (var y) (var z)) 
    (f6 (f82 (var w)) (f84 (var x)) (f83 (var y))))
\end{lstlisting}
\end{minipage}
\end{framed}
\caption{Constraint system generated from \protect\hosa.}\label{fig:hosa:constraints}
\vspace{2mm}
\end{subfigure}

\begin{subfigure}[b]{\linewidth}
\begin{framed}
\begin{minipage}{0.22\linewidth}
\begin{lstlisting}[emph={[2] f1,f2,f3,f4,f5,f6,f71,f72,f78,f71,f72,f73,f74,f75,f76,f77,f78,f79,f80,f81,f82,f83,f84,f85,f86,f87},style=cs,emph={x,y,z,w}]
f1(x,y) = x + y
f2(x,y,z) = x + y
f3(x,y,z) = z
f4(x,y) = x + y
f5(x,y,z) = x + y
f6(x,y,z) = z
\end{lstlisting}
\end{minipage}
\begin{minipage}{0.22\linewidth}
\begin{lstlisting}[emph={[2] f1,f2,f3,f4,f5,f6,f71,f72,f78,f71,f72,f73,f74,f75,f76,f77,f78,f79,f80,f81,f82,f83,f84,f85,f86,f87},style=cs,emph={x,y,z,w}]
f71(x) = x
f72(x) = x
f73(x) = x
f74(x) = x
f75(x) = x
f76(x) = x
\end{lstlisting}
\end{minipage}
\begin{minipage}{0.22\linewidth}
\begin{lstlisting}[emph={[2] f1,f2,f3,f4,f5,f6,f71,f72,f78,f71,f72,f73,f74,f75,f76,f77,f78,f79,f80,f81,f82,f83,f84,f85,f86,f87},style=cs,emph={x,y,z,w}]
f77(x) = x
f78(x,y) = x + y
f79(x) = x
f80(x,y,z,w) = x + y
f81(x) = x
f82(x) = x
\end{lstlisting}
\end{minipage}
\begin{minipage}{0.22\linewidth}
\begin{lstlisting}[emph={[2] f1,f2,f3,f4,f5,f6,f71,f72,f78,f71,f72,f73,f74,f75,f76,f77,f78,f79,f80,f81,f82,f83,f84,f85,f86,f87},style=cs,emph={x,y,z,w}]
f83(x) = x
f84(x) = x
f85(x,y,z,w) = z
f86 = 0
\end{lstlisting}
~\\
~
\end{minipage}
\end{framed}
\caption{Model inferred by \protect\gubs\ on the generated constraints.}\label{fig:hosa:model}
\vspace{2mm}
\end{subfigure}

\begin{subfigure}[b]{\linewidth}
\begin{framed}
\begin{minipage}{\linewidth}
\centering
\begin{lstlisting}[style=ocaml,basicstyle=\scriptsize\ttfamily,emph={i,j,k,l,ijk,ij},emph={[2] append,map,prependAll,f1,f3,f2,f5,f4,f6}]
       map : $\forall$ijk.($\forall$ l.L[l](a) -> L[l+i](a)) ->  L[k](L[j](a)) -> L[k](L[i+j](a))
    append : $\forall$ij.L[i](a) -> L[j](a) -> L[i+j](a)
prependAll : $\forall$ijk.L[i](a) ->  L[k](L[j](a)) -> L[k](L[i+j](a))
\end{lstlisting}
\end{minipage}
\end{framed}
\caption{Inferred size type obtained by instantiating the template types with the model computed by \protect\gubs.}\label{fig:hosa:inferred}
\end{subfigure}
\vspace{-7mm}
\caption{Sized-type inference carried out by \protect\hosa\ on
  {\lstinline[style=ocaml,emph={xs},emph={[2]append,prependAll,map}]!prependAll xs ys = map (append xs) ys!}.}
\label{fig:hosa}
\end{figure}

Various successful approaches to automatic verification of termination
properties of higher-order functional programs are based on \emph{sized-types}~\cite{HPS:POPL:96}.
Here, a type carries not only some information about the \emph{kind} of each
object, but also about its \emph{size}, hence the name.  This
information is then exploited when requiring that recursive calls are
done on arguments of \emph{strictly smaller} size.

In recent work~\cite{ADL:EV:17} we have taken a fresh look at sized-types, 
with particular emphasis towards application in runtime analysis and automation. 
A result of this work is the tool \hosa, which given a program written in a simple 
typed higher-order language, annotates all types with size information. 
\gubs\ is a central ingredient of \hosa.

Consider the function, 
\lstinline[style=ocaml,emph={xs},emph={[2]append,prependAll,map}]!prependAll xs ys = map (append xs) ys!, 
which prepends a given list $xs$ to all elements of it second argument $ys$, 
itself a list of lists.
Sized-type inference with \hosa\ works by first decorating 
datatypes occurring in the types of functions with size indices, 
resulting in so called \emph{template types}, see \Cref{fig:hosa:templates}.
Size indices are annotated here in square brackets. These template types make reference to so far undetermined functions.
Inference then amounts simply to type-checking, where however, type-checking emits constraints 
relating the different size indices. The emitted constraint system 
is depicted in \Cref{fig:hosa:constraints} in S-expression notation, the notation expected by \gubs.  
A model found by \gubs\ on the resulting constraint system, such as the one depicted in \Cref{fig:hosa:model}, 
is then used to construct concrete sized-types from the initial template types, see \Cref{fig:hosa:inferred}.

It is worthy of note that in this example, the inferred types are precise and thus informative, not least, 
because of the minimisation techniques incorporated in \gubs, as outlined in Section~\ref{s:implementation}.
Also worthy of note, including the time spend by \gubs, the depicted sized-types were found in a quarter of a second, 
on one of the authors' laptops. 

\hosa\ is also capable of instrumenting a given program, by threading through the computation a program counter.
This way, the runtime of a program is reflected in its sized-type and thus \hosa\ can give quantitative information
on the runtime of programs. 
\hosa\ is able to analyse the runtime of a series of examples, fully automatically, which
cannot be handled by most competitor methodologies (see e.g.,~\cite{HDW:POPL:17}). 
Noteworthy, \hosa\ can deal with a variety of examples
whose runtime is not linear, e.g., sorting algorithms and non-trivial list functions. \gubs\ is capable of 
dealing with reasonably sized constraint systems in this context. For instance, the runtime analysis of 
the function which computes the cross-product of two lists in quadratic time, itself defined in terms of two folds, 
relies on 87 functions related by 85 constraints. To this end, \gubs\ analyses $10$ SCCs individually. 
The computed bounds are tight, 
the overall procedure takes just three quarters of a second on this example. 

\paragraph*{Synthesis of Polynomial Interpretation.}

\begin{figure}[t]
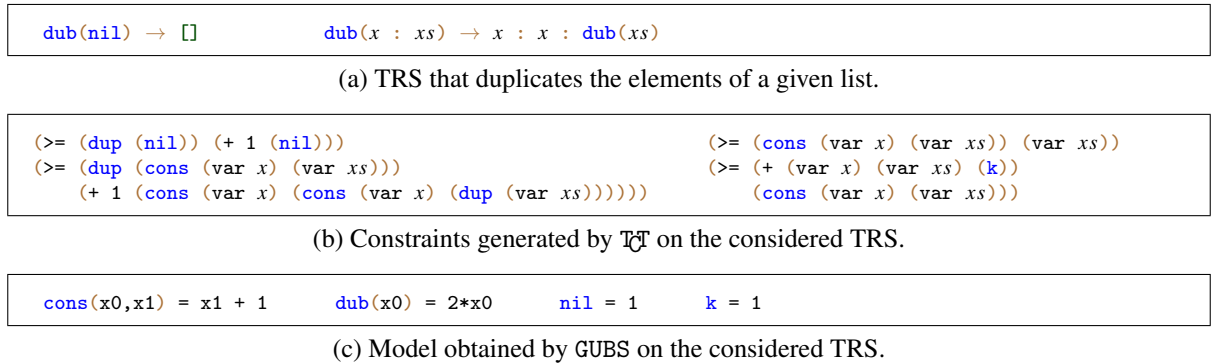

\begin{subfigure}[b]{\textwidth}
\begin{framed}
  \begin{minipage}{\linewidth}
\centering
\begin{lstlisting}[style=ocaml,basicstyle=\scriptsize\ttfamily,emph={x,xs},emph={[2] dub,nil,[],:}]
 dub(nil) -> []           dub(x : xs) -> x : x : dub(xs)
\end{lstlisting}
\end{minipage}
\end{framed}
\caption{TRS that duplicates the elements of a given list.}\label{fig:trs:input}
\vspace{2mm}
\end{subfigure}

\begin{subfigure}[t]{\linewidth}
\begin{framed}
\begin{minipage}{\linewidth}
\begin{lstlisting}[emph={[2] cons,dup,nil,k},style=cs,basicstyle=\scriptsize\ttfamily,emph={x,xs,x1,x2y,z,w}]
(>= (dup (nil)) (+ 1 (nil)))                                (>= (cons (var x) (var xs)) (var xs))
(>= (dup (cons (var x) (var xs)))                           (>= (+ (var x) (var xs) (k)) 
    (+ 1 (cons (var x) (cons (var x) (dup (var xs))))))         (cons (var x) (var xs)))
\end{lstlisting}
\end{minipage}
\end{framed}
\caption{Constraints generated by \protect\tct\ on the considered TRS.}\label{fig:trs:constraints}
\vspace{2mm}
\end{subfigure}

\begin{subfigure}[b]{\linewidth}
\begin{framed}
\begin{minipage}{\linewidth}
\begin{lstlisting}[emph={[2] cons,dub,nil,k},style=cs,basicstyle=\scriptsize\ttfamily,emph={x,y,z,w}]
 cons(x0,x1) = x1 + 1      dub(x0) = 2*x0      nil = 1      k = 1
\end{lstlisting}
\end{minipage}
\end{framed}
\caption{Model obtained by \protect\gubs\ on the considered TRS.}\label{fig:trs:model}
\vspace{2mm}
\end{subfigure}
\vspace{-8mm}
\caption{Synthesis of polynomial interpretations in \protect\tct.}
\label{fig:trs}
\end{figure}

\emph{Term rewriting systems} (\emph{TRSs} for short) provide an abstract model of computation that is at the heart of functional and declarative programming. 
In the last decade termination and resource behaviour of TRSs have been investigated actively. \emph{Polynomial interpretation} is an elementary method in this context.
In its most simple setting, a polynomial interpretation $\II$, i.e.\ interpretation over polynomials as above, \emph{orients} a TRS $\RS$
if $\im[\alpha]{l} >_\N \im[\alpha]{r}$ holds for all rules $l \to r \in \RS$. Together with certain monotonicity constraints on the interpretation, 
orientability implies termination. When the interpretation of constructors are additionally sufficiently constrained, 
quantitative information on the runtime of $\RS$ can be obtained. 
Notably, orientation constraints and monotonicity constraints are straight forward translated into a constraint system. 
Consider for instance the TRS consisting of the two rules depicted in \Cref{fig:trs:input}, 
corresponding constraints are given in \Cref{fig:trs:constraints}.

% In \Cref{fig:trs:model}, symbols \ttt{dub}, \ttt{cons} and \ttt{nil}
% denote the interpretation of the corresponding functions from the input rewrite system. 
The two constraints on the left enforce the orientation.
The first constraint on the right enforces sufficient monotonicity constraints on the interpretation, viz, \ttt{cons} should be monotone in its second argument. 
The final rule enforces that \ttt{cons} is interpreted by a strongly linear interpretation $\inter{cons}(x,xs) \leqslant x + xs + \ttt{k}$ for some constant $\ttt{k} \in \N$, thereby 
relating the interpretation of lists linearly to its size. 
The so computed interpretation is depicted in \Cref{fig:trs:model} and witnesses that the runtime complexity, i.e., the number of reduction steps as measured in the size of the input list, 
is linear. 

So far, we did not conduct a thorough investigation of the strength of \gubs\ in this context. We expect however an increase in strength and execution time
due to the incorporation of \gubs\ in \tct.

%%% Local Variables:
%%% mode: latex
%%% TeX-master: "paper"
%%% End:

\section{Experimental Evaluation}
\label{s:evaluation}

We conducted preliminary experiments to show the viability of \gubs{} and the implemented methods.
The considered examples are the constraints generated from \hosa{} for the time and size complexity analysis of a set of functional programs.
We performed the experiment with three different strategies:
$(i)$ \ttt{inc} repeatedly tries to synthesize a model using abstract templates of an increasing degree up to degree four;
$(ii)$ \ttt{inc+simp} additionally applies simplification and minimisation;
$(iii)$ \ttt{inc+simp+scc} additionally performs the SCC decomposition.

\begin{table}[htpb]
  \centering
  \begin{tabular}{c|c|c|c}
                  & \ttt{inc} & \ttt{inc+simp} & \ttt{inc+simp+scc} \\
    \hline
    \ttt{SAT}(1)  & 17        & 17             & 18                 \\
    \ttt{SAT}(2)  & 2         & 4              & 3                  \\
    \ttt{SAT}(3)  & 0         & 0              & 1                  \\
    \ttt{Timeout} & 3         & 1              & 0                  \\
    % \hline
    % Average time    & 17.16           & 4.59                          & 4,95 
  \end{tabular}
  \caption{Experimental evaluation conducted with \gubs.}
  \label{tab:eval}
\end{table}

\Cref{tab:eval} illustrates the summary of the conducted experiment.
The details of the experiment are available online\footnote{See \url{http://cbr.uibk.ac.at/tools/gubs/experiments}}.
\ttt{SAT}($n$) indicates that a model could be successfully inferred where the maximal degree of the model is $n$, which corresponds in this case to the asymptotic worst-case bounds of the original problem and \ttt{Timeout} indicates that the computation timed-out after $90$ seconds.
The strategy \ttt{inc} conceptually corresponds to the current strategy applied in \tct{} to synthesize polynomial interpretations.
We see that the experiments improve with syntactical Simplifications turned on.
Furthermore, the full strategy is the only one that finds a model for the \ttt{insertionsort} example.
% Most of the example fraction of a second, whereas \ttt{insertionsort} needs around $13$ seconds.

In particular in the context of rewriting several tools have been established that support constraint solving of diophantine constraints and the synthesis of models of constraint systems.
Consider for example,
% \tytt\footnote{See \url{http://cl-informatik.uibk.ac.at/software/ttt2/}},
% \cime\footnote{See \url{http://cime.lri.fr/}},
% \satchmo\footnote{See \url{https://github.com/jwaldmann/satchmo}} and
% \aprove\footnote{See \url{http://aprove.informatik.rwth-aachen.de/}}.
\tytt, \cime, \satchmo\ and \aprove\footnote{See \url{http://cl-informatik.uibk.ac.at/software/ttt2}, \url{http://cime.lri.fr}, \url{https://github.com/jwaldmann/satchmo} and \url{http://aprove.informatik.rwth-aachen.de} respectively}.
However, without further consideration a direct comparison is difficult.
None of the aforementioned tools can handle the given example systems directly, as the tools are usually tuned for a more specific setting.

\section{Conclusion and Future Work}

We have described \gubs, an open source tool for synthesising functions over the naturals that satisfy a given set of inequalities.
The development of \gubs\ was motivated by the lack of dedicated constraint solvers for polynomial inequalities. 
% To this end, \gubs\ heavily relies on SMT-solvers, various simplification techniques and a separate SCC analysis.
% Nowadays, \gubs\ forms the back-bone of \tct\ and \hosa, two static analysis tools. 

In future work, we would like to extend \gubs\ in various directions. 
In the short term, we would like to improve the methods that are currently implemented. 
This includes dedicated synthesis techniques for certain subclasses of constraint systems or models, e.g., 
via reductions to \emph{linear programming}. It also includes the search
for suitable divide-and-conquer methods. 
Model synthesis is in general not modular, however for certain classes, e.g. disjoint systems, synthesis is modular. 
The aforementioned SCC analysis is a first step into this direction. 
Another direction for future work is to extend upon the set of models that can be found in \gubs.
Currently, \gubs\ is only able to synthesise weakly monotone functions over the naturals. It would be interesting, 
for instance, to include support of integers and rationals, together with non-monotone functions such as subtraction and division. 
This is clearly feasible with the current toolset underlying \gubs.
% Related, an extension of \gubs\ to an SMT solver for dealing with non-linear integer arithmetic could also be worthwhile. 
Finally, and maybe most importantly, we would like to see further applications of our tool. 

%%% Local Variables:
%%% mode: latex
%%% TeX-master: "paper"
%%% End:

%%%%%%%%%%%%%%%%%%%%%%%%%%%%%%%%%%%%%%%%%%%%%%%%%%%%%%%%%%%%%%%%%%%%%%%%%%%%%%

% \nocite{*}
\bibliographystyle{eptcs}
\bibliography{references}
\end{document}